\documentclass[preprint,amsmath,amssymb,nofootinbib,superscriptaddress]{revtex4}

\usepackage{epsfig}
\usepackage[utf8]{inputenc}
\usepackage{color}
\usepackage{epstopdf}
\usepackage{multirow}

\newcommand{\be}{\begin{eqnarray}}
\newcommand{\ee}{\end{eqnarray}}

\newcommand{\X}{\ensuremath{X(3872)}}

\newcommand{\itp}{\affiliation{CAS Key Laboratory of Theoretical Physics, Institute of Theoretical Physics,\\ Chinese Academy of Sciences, Beijing 100190, China}}

\newcommand{\ucas}{\affiliation{University of Chinese Academy of Sciences, Beijing 100049, China}}

\newcommand{\uestc}{\affiliation{School of Physics, University of Electronic Science and
Technology of China,\\ Chengdu 610054, China}} 

\newcommand{\imp}{\affiliation{Institute of Modern Physics, Chinese Academy of Sciences, Lanzhou 730000, China}}

\begin{document}

\title{Semi-inclusive lepto-production of hidden-charm exotic hadrons}

\author{Zhi Yang}
\email{zhiyang@uestc.edu.cn}
\uestc
\imp

\author{Feng-Kun Guo}
\email{fkguo@itp.ac.cn}
\itp
\ucas

\date{\today}

\begin{abstract}
We investigate the semi-inclusive production of hidden-charm exotic states, including $X(3872)$, $Z_c$, $Z_{cs}$ and the pentaquark $P_c$ states, in lepton-proton scattering processes. These hadrons are close to the thresholds of a pair of charm and anticharm hadrons, and are assumed to have hadronic molecular structure as their main components.
In order to give order-of-magnitude estimates of the cross sections, we use Pythia to simulate the short-distance productions of the constituent hadrons, which then rescatter to form the exotic hadrons. 
The estimates for the $X(3872)$ and $Z_c(3900)^\pm$ are not in conflict with the upper limits measured at the COMPASS experiment for the exclusive photoproduction process.
The results here indicate that the considered hidden-charm states can be copiously produced at the proposed electron-ion colliders EicC and US-EIC.
 
\end{abstract}

\maketitle

\section{Introduction}

Since the discovery of the $X(3872)$~\cite{Choi:2003ue}, lots of  charmonium-like structures have been observed with properties beyond the expectations of conventional quark models. 
In the baryon sector, the LHCb Collaboration announced the observation of hidden-charm pentaquark states in the $J/\psi p$
invariant mass distribution in the $\Lambda_b^0\to J/\psi K^- p$ decay in 2015~\cite{Aaij:2015tga}, and the measurements were updated in 2019~\cite{Aaij:2019vzc}.
These new hidden-charm structures are widely regarded as candidates of exotic hadrons, and have been extensively studied (see recent reviews~\cite{Swanson:2006st,Zhu:2007wz,Chen:2016qju,Hosaka:2016pey,Richard:2016eis,Lebed:2016hpi,Esposito:2016noz,Guo:2017jvc,Olsen:2017bmm,Kalashnikova:2018vkv,Liu:2019zoy,Brambilla:2019esw,Guo:2019twa} for more information).

So far most of the observations of these new structures were made at electron-positron colliders and hadron colliders (including the $B$ factories). Nevertheless, there have been attempts searching for them in photoproduction processes, which are useful not only in confirming their existence but also in gaining more insights into their nature. In particular, the photoproduction processes are free of triangle singularities that sometimes induce ambiguities in interpreting the observed peaks as resonances~\cite{Wang:2013cya,Wang:2013hga,Liu:2014spa,Szczepaniak:2015eza,Guo:2015umn,Liu:2015fea, Albaladejo:2015lob, Pilloni:2016obd, Gong:2016jzb, Aaij:2019vzc} (see Ref.~\cite{Guo:2019twa} for a review). Hence, the photo-induced production of the pentaquarks has been proposed in order to discriminate their nature~\cite{Wang:2015jsa,Huang:2016tcr,Blin:2016dlf,Karliner:2015voa,Kubarovsky:2015aaa}. 
The COMPASS Collaboration searched for exclusive photoproduction of the $Z_c(3900)^\pm\to J/\psi\pi^\pm$ and no signal was found~\cite{Adolph:2014hba}. Later on, they searched for muoproduction of the $X(3872)\to J/\psi\pi^+\pi^-$ and reported the observation of a structure with a mass around 3.86~GeV~\cite{Aghasyan:2017utv}. The $C$-parity is, however, negative, opposite to that of the $X(3872)$~\cite{Aaij:2013zoa,Zyla:2020zbs}. Its relation to the $h_c(2P)$ and other $PC=++$ states around 3.9~GeV through heavy quark spin symmetry deserves to be explored. In Ref.~\cite{Dong:2021juy}, this negative $C$-parity state is suggested to be an isoscalar $D\bar D^*$ molecule with $J^{PC}=1^{+-}$ as the spin partner of the $X(3872)$. In this regard, the recently reported narrow structure $X(4014)$ with a mass of $(4014.4\pm4.1\pm0.5)$~MeV and a width of $(6\pm16\pm6)$~MeV~\cite{Wang:2021sty} is an excellent candidate for the $2^{++}$ spin partner of the $X(3872)$ that is generally predicted in hadronic molecular models~\cite{Tornqvist:1993ng,Wong:2003xk,Swanson:2005tn,Nieves:2012tt,Guo:2013sya,Albaladejo:2015dsa,Baru:2016iwj,Dong:2021juy,Chen:2021iaw}. Its mass and width nicely agree with the predictions in Ref.~\cite{Albaladejo:2015dsa}.
The GlueX Collaboration searched for the $P_c$ pentaquark states through the near-threshold $J/\psi$ exclusive photoproduction off the proton in Hall D at Jefferson Lab (JLab) and no evidence was found~\cite{Ali:2019lzf}, implying a small branching fraction for the decay into $J/\psi p$~\cite{Cao:2019kst,Winney:2019edt}. There is also a proposal to search for the $P_c$ in Hall C at JLab~\cite{Meziani:2016lhg}.

There have been lots of estimates of the photoproduction cross sections of hidden-charm states for exclusive processes~\cite{Liu:2008qx,Galata:2011bi,Lin:2013mka,Lin:2013ppa,Huang:2013mua,Adolph:2014hba,Wang:2015jsa,Wang:2015lwa,Kubarovsky:2015aaa,Karliner:2015voa,Huang:2016tcr,Blin:2016dlf,
Paryev:2018fyv,Wang:2019krd,Goncalves:2019vvo,Wu:2019adv,Winney:2019edt,Xie:2020niw,Yang:2020eye,Albaladejo:2020tzt}.
The analysis of exclusive productions of pentaquarks at both JLab and the Electron-ion collider in China (EicC)~\cite{Yang:2020eye} suggests that the colliding mode is more useful than the fixed-target mode at JLab to extract the feeble exclusive pentaquark signal. 
For the three experiments of interest here, i.e., EicC, COMPASS and the Electron-Ion Collider in US (US-EIC), the c.m. energies are much higher than the hidden-charm hadron masses, and thus allow for their production together with many other light hadrons. Although semi-inclusive productions have larger background compared to the exclusive processes, the production rates could exceed the exclusive processes by a few orders of magnitude, which makes them promising in studying the exotic hadrons. 

In this paper, we will estimate the semi-inclusive lepto-production rates of the $X(3872)$~\cite{Choi:2003ue}, $Z_c(3900)$~\cite{Ablikim:2013mio,Liu:2013dau}, $Z_c(4020)$~\cite{Ablikim:2013wzq}, $Z_{cs}(3985)$~\cite{Ablikim:2020hsk}, and the $P_c$ states~\cite{Aaij:2015tga,Aaij:2019vzc} together with their spin partners predicted in the hadronic molecular model~\cite{Xiao:2013yca,Liu:2019tjn,Xiao:2019aya,Sakai:2019qph,Du:2019pij,Wang:2019ato,Du:2021fmf} (We regard the $Z_{cs}(4000)$ reported by LHCb as due to the same origin following the suggestion of Refs.~\cite{Yang:2020nrt,Ortega:2021enc}.). Results will be presented for three experiments: the proposed US-EIC with the $ep$ c.m. energy covering 30 $\sim$ 140~GeV and a luminosity of
$10^{34}\;\text{cm}^{-2}\text{s}^{-1}$ or higher~\cite{Accardi:2012qut}, EicC at lower energies (15$\sim$25~GeV)~\cite{CAO:2020EicC,Anderle:2021wcy}, and
the COMPASS experiment with muon beams at CERN~\cite{Abbon:2007pq}.

It is worthwhile to notice that almost all the existing literature~\cite{Liu:2008qx,Galata:2011bi,Lin:2013mka,Lin:2013ppa,Huang:2013mua,Adolph:2014hba,Wang:2015jsa,Wang:2015lwa,Kubarovsky:2015aaa,Karliner:2015voa,Huang:2016tcr,Blin:2016dlf,
Paryev:2018fyv,Wang:2019krd,Goncalves:2019vvo,Wu:2019adv,Winney:2019edt,Xie:2020niw,Yang:2020eye,Albaladejo:2020tzt} estimate the photoproduction cross sections of hidden-charm or hidden-bottom resonances using the vector-meson dominance model, where the photon is assumed to convert to a highly virtual vector heavy quarkonium, such as the $J/\psi$ and interacts with the proton.
However, Ref.~\cite{Du:2020bqj} pointed out the importance of open-charm channels in the $J/\psi$ near-threshold photoproduction, and the estimated total cross section by considering only the intermediate $\Lambda_c\bar D^{(*)}$ channels quantitatively agree with the GlueX measurement in the near-threshold region~\cite{Ali:2019lzf}.
The reliability of the vector-meson dominance model in the photoproduction of vector heavy quarkonia is also questioned recently based on Dyson--Schwinger equation calculations in Ref.~\cite{Xu:2021mju}.
Here we will consider the mechanism of open-charm channels---the charmed-hadron pairs are produced first and the hadronic molecular candidates are produced through their final state interactions.

This paper is organized as follows. We will first introduce the formalism of the semi-inclusive production of hadronic molecules in Sec.~\ref{sec:anaframe}. Then we present the production rates of some typical exotic hadrons, including $X(3872)$, $Z_c(3900)$, $Z_c(4020)$, $Z_{cs}(3985)$, and seven $P_c$ states, which are considered as hadronic molecules of $D\bar D^*+c.c.$, $D\bar D^*+c.c.$, $D^*\bar D^*$, $D_s\bar D^*+ D_s^*\bar D$, and $\Sigma_c^{(*)}\bar D^{(*)}$,  respectively, at the proposed EicC, US-EIC and the currently running COMPASS at Sec.~\ref{sec:numresults}. The last section is a short summary.

\section{Cross section estimate}
\label{sec:anaframe}

\begin{figure*}[tbp]
\centering
\includegraphics[width=0.4\textwidth]{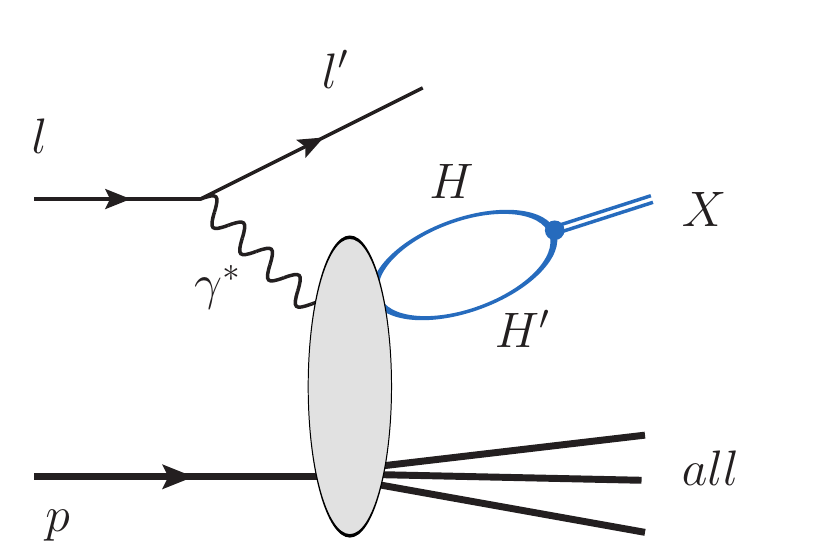}
\caption{\label{fig:diagram} 
The production mechanism considered for the hadronic molecule production, where $X$ denotes a $HH^{\prime}$ molecular state, in the lepton-proton semi-inclusive reaction. Here $all$
represents all the produced particles other than the final lepton and $X$.}
\end{figure*}

A resonance corresponds to a pole of the $S$-matrix. For the $S$-matrix element between the initial state, which is a lepton-proton pair here, and the final state containing the exotic hadron of interest, one may insert a complete set of states that can couple to the exotic hadron. All of them can contribute to the production. Yet, it is natural to expect the ones that couple most strongly to the exotic hadron have the most significant contributions.
For hadronic molecules, such states correspond to the particles that are bound together to give the dominant part of the wave function at large distances.

The lepto-production of an exotic hadron $X$ in the molecular picture is shown in Fig.~\ref{fig:diagram},
which assumes the mechanism that the formation of a hadronic molecule occurs after the production of its constituents.
Details of the mechanism can be found in Ref.~\cite{Artoisenet:2010uu,Guo:2014ppa,Guo:2014sca}. The mechanism can be applied to the $X(3872)$ and $Z_c(3900)$ which are known to couple strongly to the $D\bar D^*$ pair (and their antiparticles)~\cite{Gokhroo:2006bt,Aubert:2007rva,Adachi:2008sua,Ablikim:2013xfr,BESIII:2013qmu,Zyla:2020zbs} as well as to the $P_c$ hidden-charm pentaquarks.
It has been used in estimating the prompt production cross sections of the $X(3872)$~\cite{Artoisenet:2009wk,Albaladejo:2017blx} and its partners~\cite{Guo:2014sca}, the $Z_c$ and $Z_b$ states~\cite{Guo:2013ufa}, the charm-strange hadronic molecules~\cite{Guo:2014ppa}, and the $P_c$ states~\cite{Ling:2021sld} at hadron colliders.
As for the production of the $X(3872)$, despite the debates~\cite{Bignamini:2009sk,Artoisenet:2009wk,Meng:2013gga,Guo:2014sca,Albaladejo:2017blx,Guo:2017jvc,Esposito:2017qef,Wang:2017gay,Braaten:2018eov,Butenschoen:2019npa,Zhang:2020dwn,Esposito:2020ywk,Braaten:2020iqw}, the estimated total cross sections~\cite{Guo:2014sca,Albaladejo:2017blx} are consistent with the CDF~\cite{Bauer:2004bc} and CMS~\cite{Chatrchyan:2013cld} measurements.

The prompt production of hadronic molecules can be factorized into short-distance and long-distance contributions~\cite{Artoisenet:2009wk,Artoisenet:2010uu}. The hadron pairs, denoted as $HH'$, are produced at short distances, and then form loosely bound hadronic molecules at long distances. The production cross section may be written as~\cite{Guo:2014sca}
\begin{eqnarray}
	\sigma[X] &=&   \frac{1}{4m_Hm_{H'}}
	 g^2 |G(E)|^2
\frac{d\sigma[HH^{\prime}(k)]}{dk}\frac{4\pi^{2}\mu}{k^{2}},
\label{eq:cross-section-final}
\end{eqnarray}
where $\mu$ is the reduced mass of the $HH^{\prime}$ pair and
$d\sigma[HH^{\prime}(k)]/dk$ is the differential
cross section of the $HH^{\prime}$ production with $k$ the three-momentum magnitude in the center-of-mass frame of
the $HH^{\prime}$ system. Without considering the final state interaction, its behavior follows the phase space,
\begin{eqnarray}
\frac{d\sigma[HH^{\prime}(k)]}{dk} \propto k^2.  \label{eq:leading_order_behaviour}
\end{eqnarray}
The exact behavior may be approximated by Monte Carlo (MC) event generators~\cite{Artoisenet:2010uu}.
The ultraviolet divergent loop function $G(k)$ can be regularized by employing a Guassian
regulator, and yields~\cite{Nieves:2012tt}
\begin{eqnarray}
	G(E) &=&  -\frac{\mu}{(2\pi)^2}\left\{\sqrt{2\pi}\,\Lambda + 2k\,\pi\,e^{-2\,k^2/\Lambda^2}\left[i-\text{erfi}\left(\frac{\sqrt{2}k}{\Lambda}\right)\right]\right\},
\label{eq:loopintegral}
\end{eqnarray}
where $k^2=2\mu(E-m_H-m_{H'})$ with $E$ the invariant mass of the $HH'$ pair,
$\text{erfi}(z)=(2/\sqrt{\pi})\int_0^z e^{t^2} dt$ is the imaginary error function and $\Lambda$ is the cut-off momentum in the Gaussian regulator. A reasonable choice of $\Lambda$ is $0.5-1$ GeV, which is used in the literature of hadronic molecular models for the states of interest here~\cite{Guo:2013sya,Liu:2019tjn,Yang:2020nrt}. The long-distance contribution is contained in the effective coupling of the hadronic molecule to its constituents $g$, whose square is given by the 
residue of the transition matrix element for the scattering $HH'\to HH'$ at the pole,
\begin{equation}
    g^2 = \lim_{E\to E_\text{pole} } \left(E^2-E^2_\text{pole} \right)\, T(E),
\end{equation}
where $E_\text{pole}=M$ for a bound or virtual state pole, and $E_\text{pole}=M-i\;\Gamma/2$ for a resonance pole,
with $M$ and $\Gamma$ the mass and width of the hadronic molecule.
In the above equation, $T(E)$ should take its value on the specific Riemann sheet where the pole is located.

For the $T$-matrix element, we consider the Lippmann-Schwinger equation,
\begin{equation}
T(E) = V + V G(E) T(E),
\end{equation}
with constant contact terms for the $HH'\to HH'$ scattering collected as the $V$ matrix, which is the leading order nonrelativistic approximation for molecular states close to thresholds.
The molecular states correspond to poles of the $T$-matrix at the solutions of  ${\rm det}[1-V\,G(E_\text{pole})]=0$.

The values of the constant contact terms for the $Z_c$ and $Z_{cs}$ states can be found in Eq.~(16) of Ref.~\cite{Yang:2020nrt}.
For the $X(3872)$ state, we take the $V$-matrix as given in Ref.~\cite{HidalgoDuque:2012pq},
\begin{equation}       
V =
\frac{1}{2}\begin{pmatrix}                
    C_0 + C_1 & C_0 - C_1 \\ 
    C_0 - C_1 & C_0 + C_1 \\ 
  \end{pmatrix},            
\end{equation}
where the two channels correspond to the neutral and charged $D\bar D^*+c.c.$ pairs.
The two low-energy parameters $C_0$ and $C_1$ can be solved by using as inputs the $X(3872)$ mass
and the isospin violation ratio of the amplitudes for the $X(3872)$ decays to $J/\psi\pi\pi$ and 
$J/\psi\pi\pi\pi$~\cite{Hanhart:2011tn,HidalgoDuque:2012pq}. While for the $P_c$ states, the relevant hadron pairs are $\Sigma_c^{(*)}\bar D^{(*)}$ that can form seven $S$-wave combinations. Heavy quark spin symmetry (HQSS) constrains that there are two low-energy parameters at leading order of the nonrelativistic expansion.
The linear combinations of the two parameters for all seven $\Sigma_c^{(*)}\bar D^{(*)}$ can be found in Refs.~\cite{Sakai:2019qph,Du:2019pij}. 
They can be fixed from reproducing the masses of two states among the $P_c(4312)$ (as the $\Sigma_c\bar D$ bound state with $J^P=1/2^-$), $P_c(4440)$ and $P_c(4457)$ (as $\Sigma_c\bar D^*$ bound states with one of them being $J^P=1/2^-$ and and the other being $3/2^-$)~\cite{Liu:2019tjn}. 
Recently, Refs.~\cite{Du:2019pij,Du:2021fmf} show that the $J/\psi p$ line shape measured by LHCb can be well described in the hadronic molecular picture with HQSS built in. Here we use the seven $P_c$ masses from the preferred scheme~II in the analysis of Ref.~\cite{Du:2019pij} as inputs to determine the two low-energy parameters in the contact terms. 

\section{Results and discussions}
\label{sec:numresults}

\begin{figure}[pth]
\centering
\includegraphics[width=0.495\textwidth]{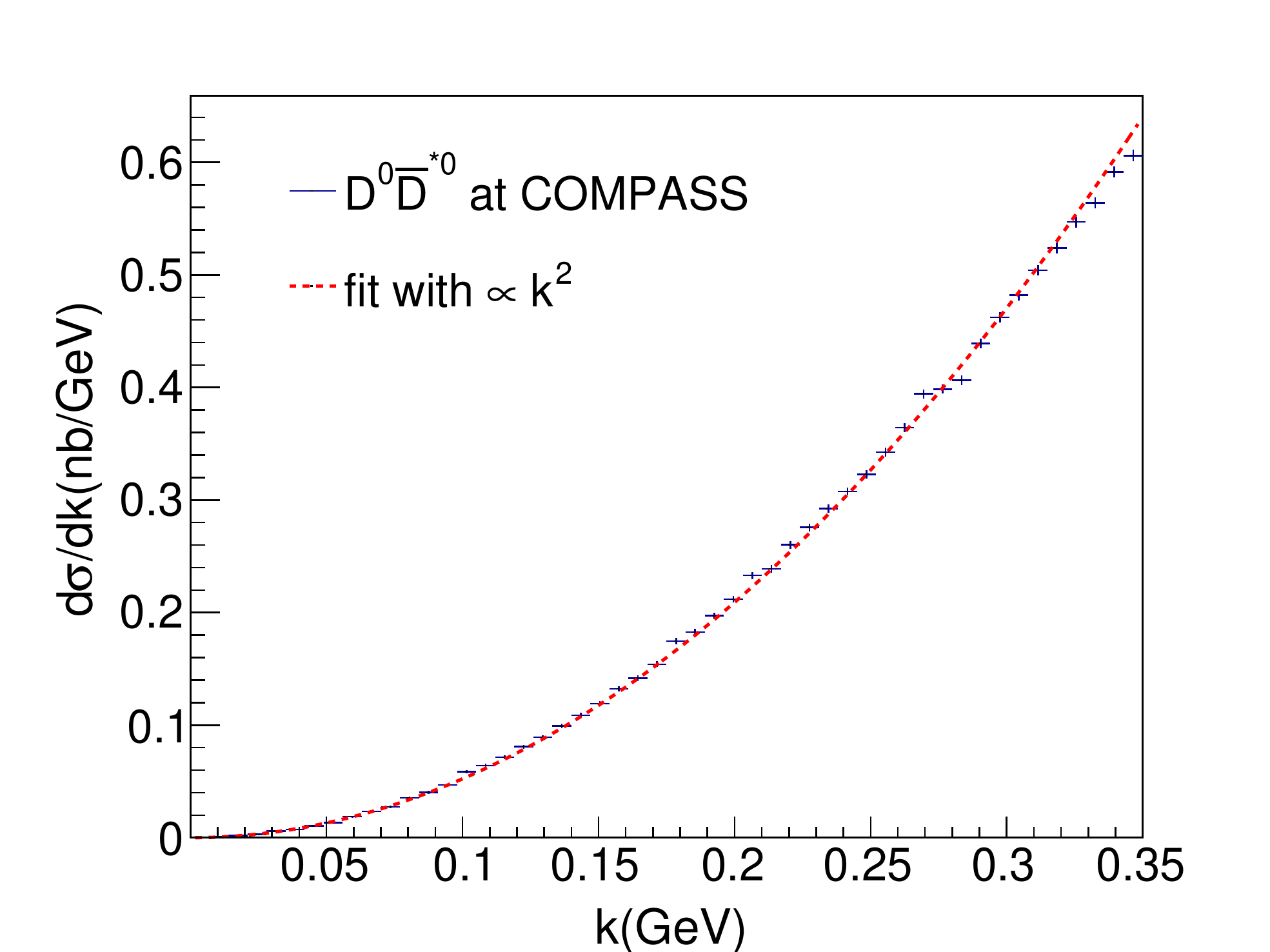}\hfill
\includegraphics[width=0.495\textwidth]{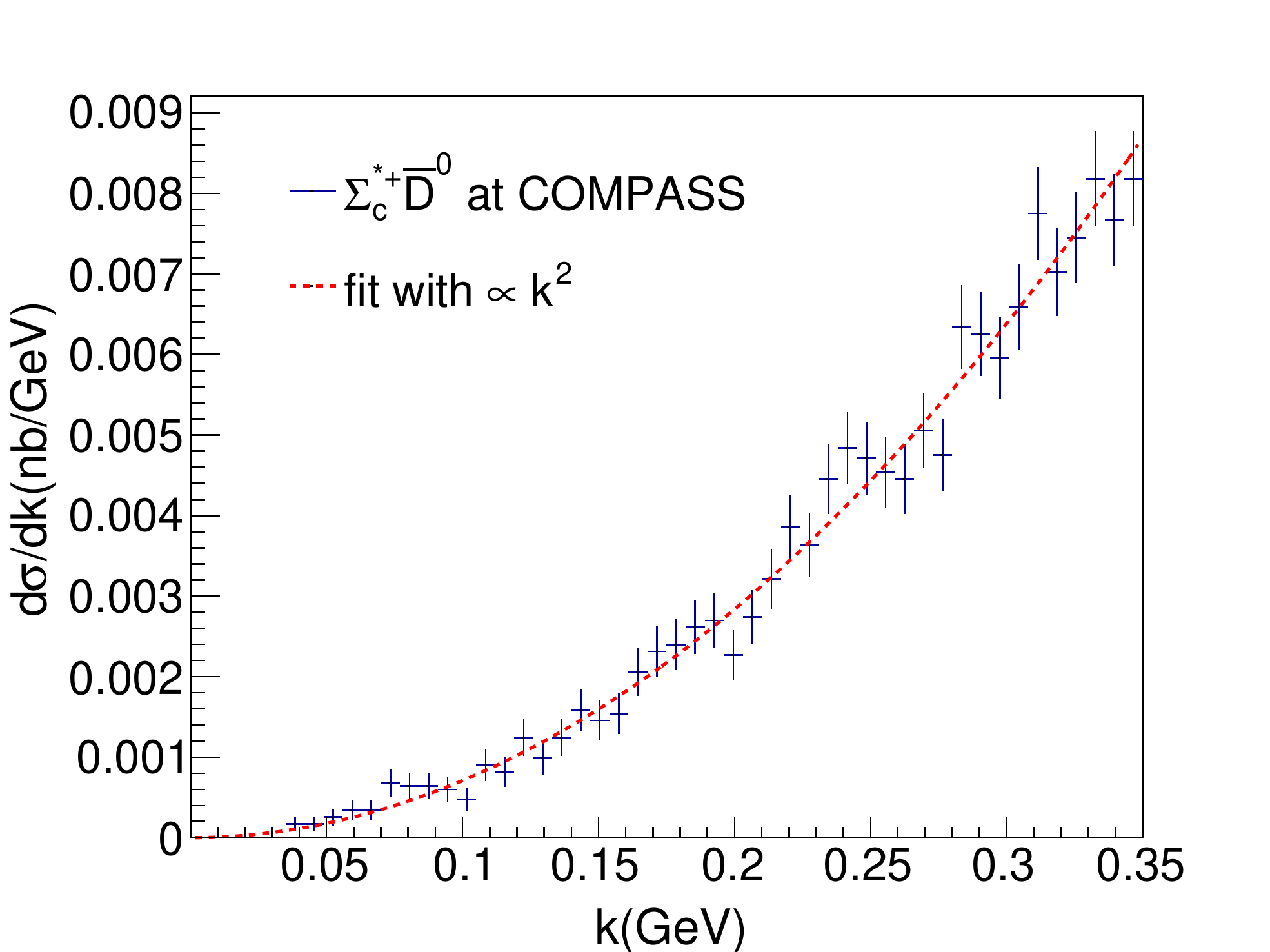} \\
\includegraphics[width=0.495\textwidth]{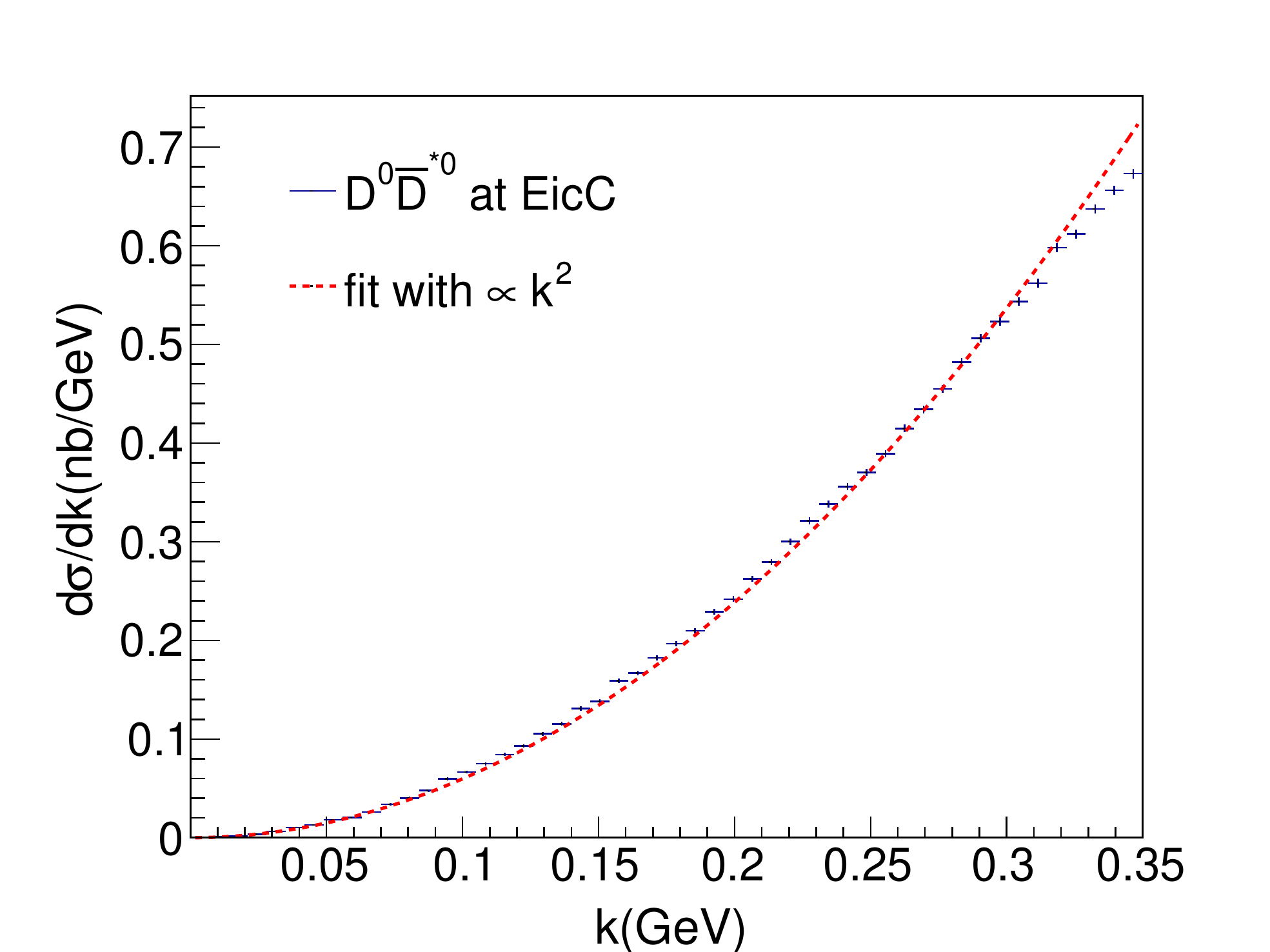}\hfill
\includegraphics[width=0.495\textwidth]{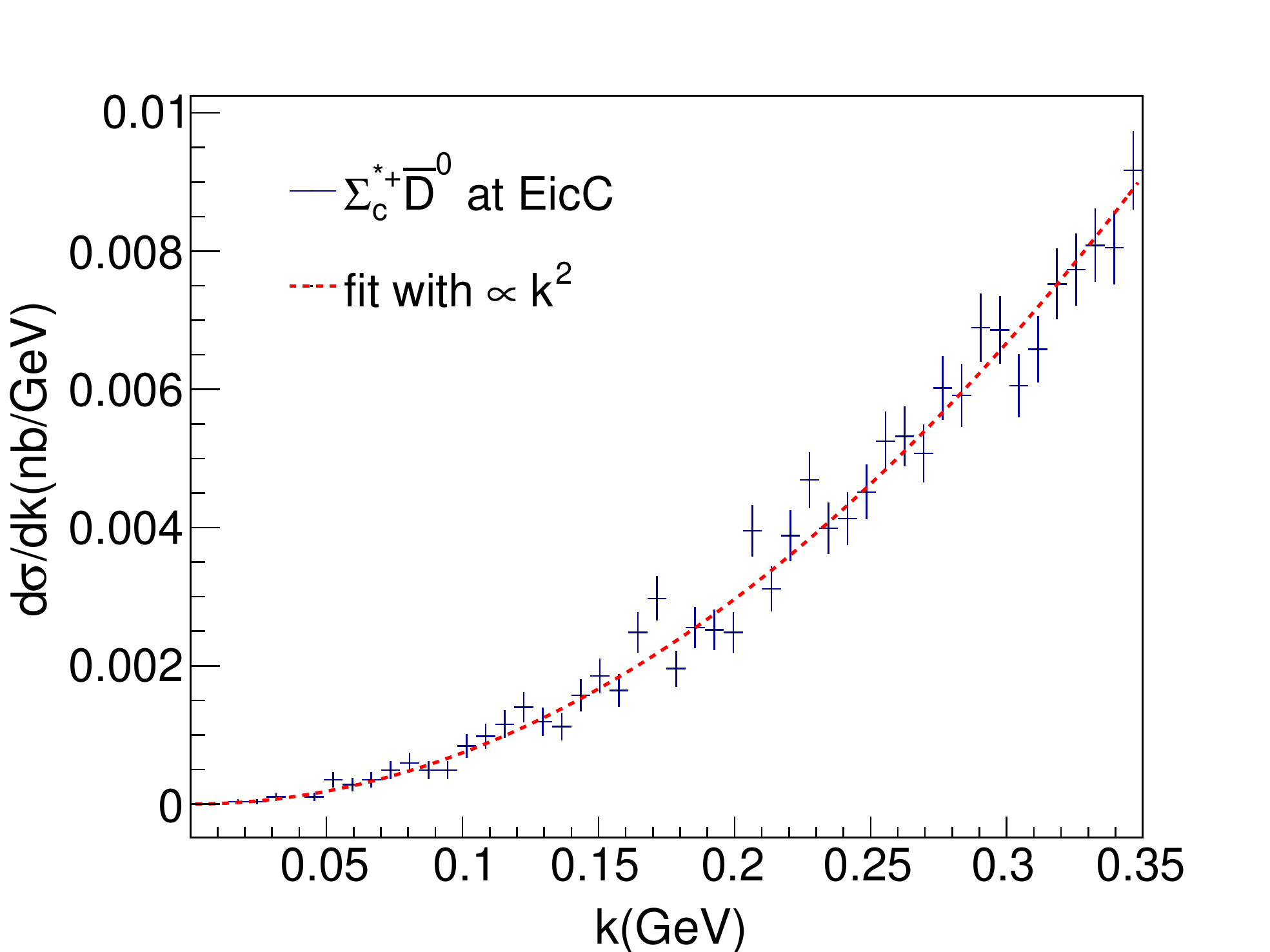} \\
\includegraphics[width=0.495\textwidth]{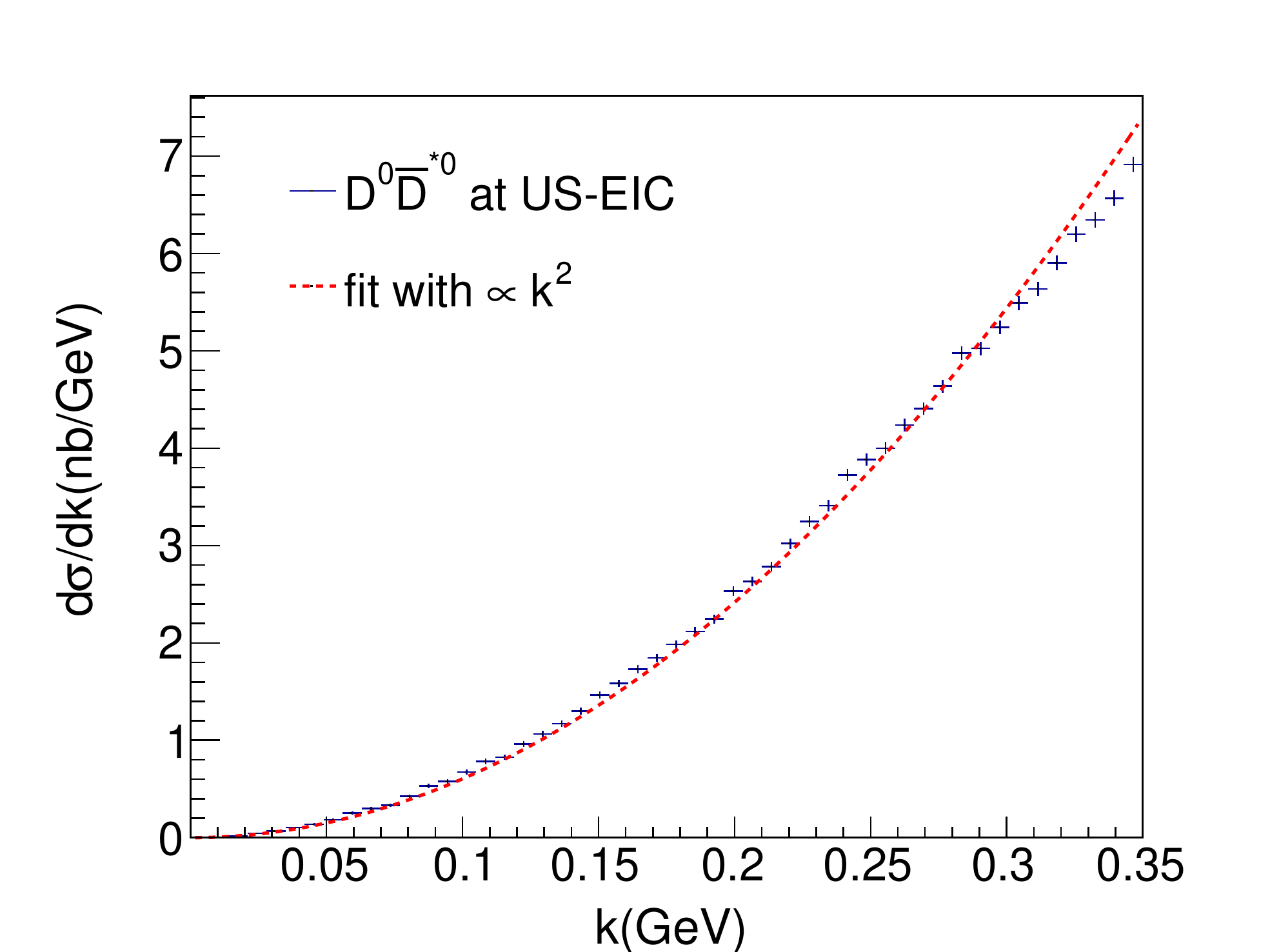}\hfill
\includegraphics[width=0.495\textwidth]{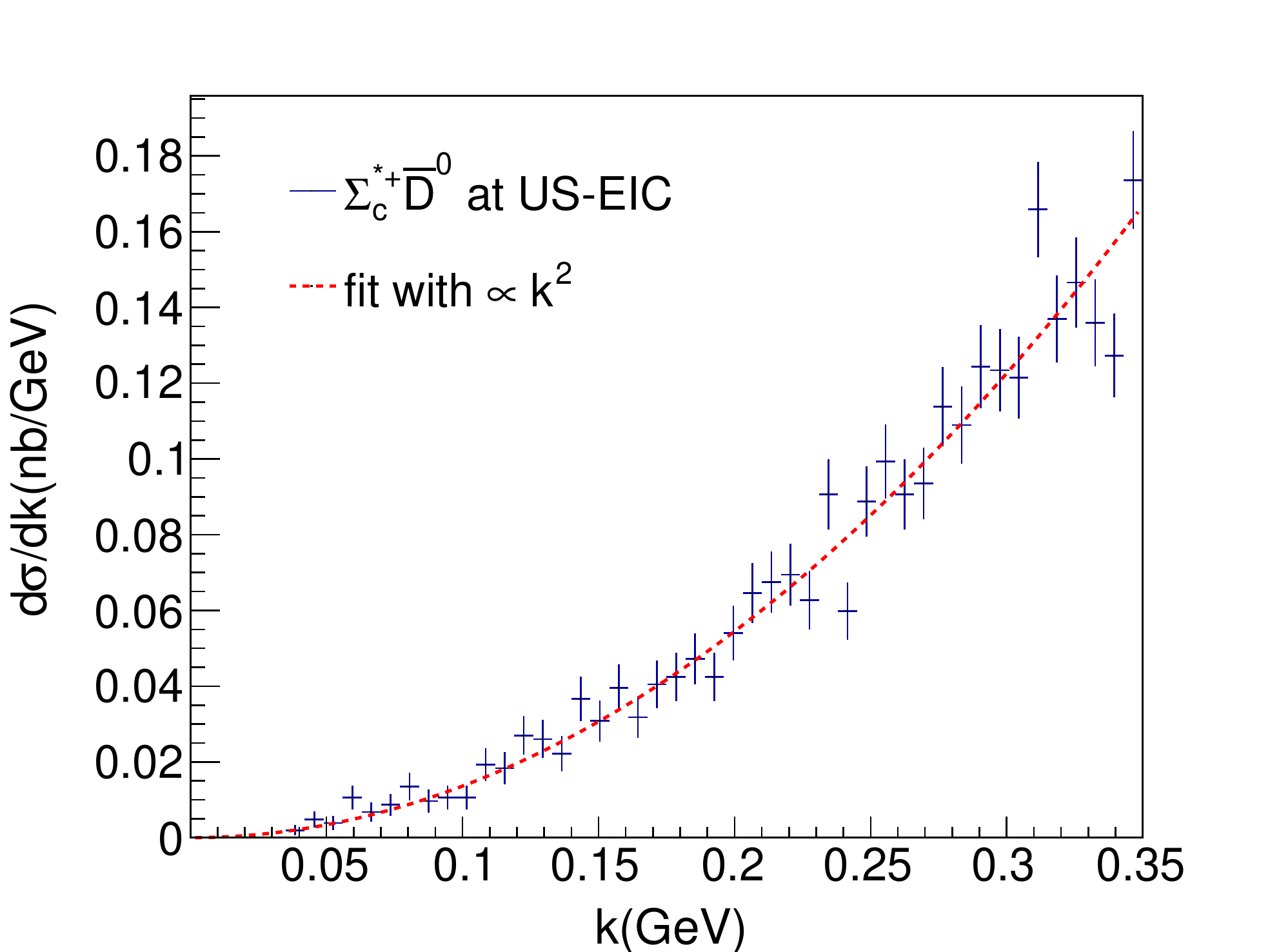} 
\caption{\label{fig:diff} 
Differential cross sections ${d\sigma}/{ dk}$ (in units of nb/GeV) 
for the semi-inclusive production $D^{0}\bar{D}^{*0}$ and $\Sigma_c^{*+}\bar{D}^0$ at COMPASS, EicC and EIC 
through lepton-proton scattering. The histograms are obtained from the Pythia event generator while the
curves are  fitted according to the momentum dependence $k^2$.}
\end{figure}

With the assumed production mechanism in Sec.~\ref{sec:anaframe}, the hadronic constituents first are produced and move with a small relative momentum such that they are in a kinematic region for a molecule to be formed. We first generate
events of the semi-inclusive lepton-proton scattering process by using the Monte Carlo generator 
Pythia~\cite{Sjostrand:2006za} at  three platforms COMPASS, EicC and US-EIC according to the energy 
configurations in Table~\ref{tab:parameters}. The pair of constituent hadrons with a small relative
momentum  were picked out from the final states generated by Pythia.\footnote{Here we choose it to be less than 350~MeV, however the exact value is not essential as the range is chosen only to determine the coefficient for Eq.~\eqref{eq:leading_order_behaviour}.} Their differential cross
sections with respect to the relative momenta are shown in Fig.~\ref{fig:diff}.
Here we take the $D^0\bar{D}^{*0}$ and $\Sigma_c^{*+}\bar{D}^0$ pairs as examples; it is shown that the distributions perfectly behave as Eq.~\eqref{eq:leading_order_behaviour}, proportional to $k^2$.

\begin{table}[t]
\caption{The energy configurations and luminosities of the three machines studied in this work. The lepton is muon for COMPASS and electron for EicC and US-EIC. Notice that COMPASS is a fixed-target experiment, and a dilution factor needs to be considered for the luminosity in estimating event numbers. 
}
\label{tab:parameters}
\begin{ruledtabular}
\begin{tabular}{l|ccc}
 &  COMPASS & EicC & US-EIC   \\
\hline
Lepton energy (GeV)   &  200      & 3.5    &  20  \\
Proton energy (GeV)    &  0      & 20   &  250  \\
Luminosity ($\text{cm}^{-2}\text{s}^{-1}$)   & $2\times10^{32}$   & $2\times10^{33}$   & $10^{34}$  \\
\end{tabular} 
\end{ruledtabular} 
\end{table}

\begin{table}[t]
\caption{Integrated cross sections (in units of pb) for $l+p\to
\text{HM}$+all, where
$\text{HM}=X(3872), \;Z_c(3900)^{0/+}, \;Z_c(4020),\;\text{and seven}\; P_c \;\text{states}$. The listed quantum numbers for these states are those in the considered hadronic molecular model.
The lepton $l$ is muon for COMPASS,and is electron for EicC and US-EIC.
Results outside (inside) brackets are obtained using cutoff $\Lambda=0.5
\;\text{GeV}\;(1\;\text{GeV})$, respectively.
}
\label{tab:cross_section}
\begin{ruledtabular}
\begin{tabular}{l|clcccc}
  & Constituents & $J^{P(C)}$ & COMPASS & EicC & US-EIC   \\
\hline
$X(3872)$     & $D\bar{D}^{*}$ & $1^{++}$ & 19(78)      & 21(89)   & 216(904)  \\
$Z_c(3900)^0$ & $D\bar{D}^{*}$ & $1^{+-}$ & $0.3\times10^3$($1.2\times10^3$)      & $0.4\times10^3$($1.3\times10^3$)   & $3.8\times10^3$($14\times10^3$)  \\
$Z_c(3900)^+$ & $D^{*+}\bar{D}^0$ & $1^{+}$ & $0.2\times10^3$($0.9\times10^3$)       & $0.3\times10^3$($1.0\times10^3$)   & $2.7\times10^3$($9.9\times10^3$)  \\
$Z_c(4020)^0$   & $D^{*}\bar{D}^{*}$ & $1^{+-}$ & $0.1\times10^3$($0.5\times10^3$)       & $0.2\times10^3$($0.6\times10^3$)       & $1.7\times10^3$($6.3\times10^3$)  \\
$Z^-_{cs}$      & $D^{*0}D_s^{-}$ & $1^{+}$ & 8.3(29)    & 19(69)    & 253(901)  \\
$Z_{cs}^{*-}$  & $D^{*0}D_s^{*-}$ & $1^{+}$ & 6.2(22)    & 14(51)    & 192(679)  \\
$P_c(4312)$   & $\Sigma_c\bar{D}$ & $\frac12^{-}$ & 0.8(4.1)    & 0.8(4.1)    & 15(73)  \\
$P_c(4440)$   & $\Sigma_c\bar{D}^{*}$ & $\frac32^{-}$ & 0.6(4.3)   & 0.7(4.7)   & 11(79)  \\
$P_c(4457)$   & $\Sigma_c\bar{D}^{*}$ & $\frac12^{-}$ & 0.5(2.0)   & 0.6(2.2)   & 9.9(36)  \\
\hline
$P_c(4380)$   & $\Sigma_c^{*}\bar{D}$ & $\frac32^{-}$ & 1.6(8.0)    & 1.6(8.4)   & 30(155)  \\
$P_c(4524)$   & $\Sigma_c^{*}\bar{D}^{*}$& $\frac12^{-}$ & 0.8(3.6)    & 0.8(3.9)   & 14(67)  \\
$P_c(4518)$   & $\Sigma_c^{*}\bar{D}^{*}$& $\frac32^{-}$ & 1.2(6.6)    & 1.2(6.9)   & 22(123)  \\
$P_c(4498)$   & $\Sigma_c^{*}\bar{D}^{*}$& $\frac52^{-}$ & 1.1(9.3)    & 1.2(9.8)  & 21(173)  \\
\end{tabular}
\end{ruledtabular} 
\end{table}

From Eq.~\eqref{eq:cross-section-final}, we can estimate the cross sections, which depend on the cut-off 
parameter $\Lambda$ in the loop function. In principle the cut-off dependence should be absorbed by that of the short-distance production of the hadron pairs. Here, as an order-of-magnitude estimate of the cross sections, we simply use the hadron pair productions from Pythia and choose the cut-off parameter to be 
$\Lambda \in [0.5,1.0]$ GeV, as that used in the calculation of the spectrum in hadronic molecular models. 
The estimated cross sections with the uncertainty caused by the cut-off range are shown in 
Table~\ref{tab:cross_section} for the $X(3872)$, $Z_c(3900)$, $Z_c(4020)$, two $Z_{cs}$ states and seven $P_c$ states at COMPASS, EicC and US-EIC. We can see that COMPASS and EicC have similar cross sections, while US-EIC have one order of magnitude
larger cross sections due to its higher center-of-mass energy. 

The COMPASS Collaboration already searched for the exclusive muoproduction of $\X$ with the reaction $\mu^+ N \to \mu^+ (J/\psi\pi^+\pi^-)\pi^\pm N'$ using data with incoming muons of 160~GeV and 200~GeV momenta~\cite{Aghasyan:2017utv}. Although no signal of the $\X$ was seen, a negative $C$-parity $\tilde{X}(3872)$ with a mass of $(3860.0\pm10.4)$~MeV and width $<51$~MeV (CL=90\%) was reported with a statistical significance of $4.1\sigma$. The measured cross section is $\sigma_{\gamma N\to \tilde X(3872) N'}\times \mathcal{B}_{\tilde X(3872)\to J/\psi\pi^+\pi^-}=(71\pm28_\text{stat}\pm39_\text{syst})$~pb. 
The upper limit for the $X(3872)$ production at COMPASS was also given as $\sigma_{\gamma N\to X(3872) N'}\times \mathcal{B}_{X(3872)\to J/\psi\pi^+\pi^-}<2.9$~pb (CL=90\%).
The upper limit for the exclusive production of the $Z_c(3900)^\pm$ at COMPASS was measured to be $\sigma_{\gamma N\to Z_c(3900)^\pm N'}\times \mathcal{B}_{Z_c(3900)^\pm\to J/\psi\pi^\pm}<52$~pb (CL=90\%) ~\cite{Adolph:2014hba}.
Although the photoproduction rates should be higher than the leptoproduction ones by roughly two orders of magnitude, the estimates for the $X(3872)$ and $Z_c(3900)^\pm$  at COMPASS listed in Table~\ref{tab:cross_section} are not in conflict with these upper limits---the semi-inclusive rates should be much higher than the exclusive ones, and the $J/\psi \pi^+\pi^-$ and $J/\psi\pi^\pm$ contribute only a small part of the decay fractions of the $X(3872)$~\cite{Zyla:2020zbs} and $Z_c(3900)$~\cite{BESIII:2013qmu}, respectively.
It would be promising to investigate the exotic states at COMPASS through the semi-inclusive processes considered here.

While for $P_c$, the GlueX measurements~\cite{Ali:2019lzf} were performed with energies just above the threshold, and far below the energies considered here. Thus, a direct comparison is not possible.
Any way, our result for the semi-inclusive cross sections of the $P_c$ states at EicC are at the order of a few pb, implying a few hundreds of pb for the corresponding photoproduction cross sections. Assuming that the $P_c \to J/\psi p$ branching fractions to be at the per cent level, we expect $\sigma(\gamma p \to P_c)\times\mathcal{B}(P_c\to J/\psi p)$ to be at the order of pb at EicC, and the cross sections at GlueX would not be larger. One the one hand, this is compatible with the GlueX upper limits (a few nb~\cite{Ali:2019lzf}); on the other hand, this means that it is difficult to observe the $P_c$ states at GlueX.

The exclusive photoproduction cross sections of $Z_{cs}^{(*)-}$ were estimated to be around 500~pb at the c.m. energy of EicC using the vector meson dominant model~\cite{Cao:2020cfx}, where the results are independent on their inner structure. The exclusive leptoproduction would be roughly two orders of magnitude smaller, making it around 5 pb. It is not in conflict with our result under the molecular assumption through semi-inclusive process shown in Table~\ref{tab:cross_section}.

Considering the luminosities given in Table~\ref{tab:parameters},
these exotic states can be copiously produced at EicC and US-EIC. Taking the $\X$ as an example, there will be around 
$8\times10^3$ and $4\times10^5$ events produced per day at EicC and US-EIC, respectively. If we set the branching fractions
$\mathcal{B}(\X\to J/\psi\pi\pi)=(3.8\pm1.2)\%$, $\mathcal{B}(J/\psi\to l^+l^-)=12\%$~\cite{Zyla:2020zbs} and assuming
the detection efficiency to be $50\%$, then the reconstructed event numbers will be about 20 and 1000 per day for EicC and US-EIC, respectively. On the other hand, due to the much higher c.m. energies, the produced exotic states at US-EIC are distributed in the large rapidity range, which brings challenges on the detector design. Therefore, further simulation of the final particle distribution at US-EIC is necessary.

\section{Summary}
\label{sec:summary}

We present order-of-magnitude estimates of the semi-inclusive production rates of hidden-charm exotic hadrons through lepton-proton scattering processes.
The calculation here is based on the hadronic molecular picture with the factorization assumption, where the hadron constituents are
produced first and subsequently bound together through the final state interaction. 
Three machines are considered, including COMPASS, EicC and US-EIC. 
The cross sections for the $X(3872)$ and $Z_c(3900)^\pm$ production at COMPASS are not in conflict with the reported upper limits for the exclusive photoproduction reactions.
According to our estimates, the proposed electron-ion colliders 
EicC and US-EIC
can produce a large number of events for the near-threshold hidden-charm molecular states. 
While for the current COMPASS experiment, it is also promising to search for these exotics through the semi-inclusive process.

\begin{acknowledgments}
Z.Y. gratefully acknowledges the insightful discussions with Meng-Lin Du and Xu Cao. This work was supported in part by the Chinese Academy of Sciences (CAS) under Grants  No.~XDB34030000 and No.~QYZDB-SSW-SYS013, 
by the National Natural Science Foundation of China (NSFC) under Grants No.~11835015, No.~12047503, No.~11961141012 and No.~11847301, by the Fundamental Research Funds for the Central Universities under Grant No. 2019CDJDWL0005, and
by NSFC and the Deutsche Forschungsgemeinschaft (DFG) through the Sino-German Collaborative
Research Center ``Symmetries and the Emergence of Structure in QCD''
(NSFC Grant No.~12070131001, DFG Project-ID 196253076 -- TRR110). 
\end{acknowledgments}

\bigskip

\bibliography{molecule_production.bib}

\end{document}